# Photothermal Microswimmer Penetrate Cell Membrane with Cavitation Bubble

Binglin Zeng,[†,‡,⊥] Jialin Lai,[†,⊥] Jingyuan Chen,[†,⊥] Yaxin Huang,[†] Changjin Wu,[¶] Chao Huang,[†] Qingxin Guo,[†] Xiaofeng Li,[†] Shuai Li,[*,§] and Jinyao Tang[*,†,‖]

[†]Department of Chemistry, The University of Hong Kong, Hong Kong, China
[‡]HKU-CAS Joint Laboratory on New Materials and Department of Chemistry, Hong Kong, China
[¶]Department of Mechanical Engineering, The University of Hong Kong, Hong Kong, China
[§]College of Shipbuilding Engineering, Harbin Engineering University, Harbin 150001, China
[‖]State Key Laboratory of Synthetic Chemistry, The University of Hong Kong, Hong Kong, China
[⊥]Authors contribute equally to this work

E-mail: lishuai@hrbeu.edu.cn; jinyao@hku.hk

## Abstract

Self-propelled micromotors can efficiently convert ambient energy into mechanical motion, which is of great interest for its potential biomedical applications in delivering therapeutics noninvasively. However, navigating these micromotors through biological barriers remains a significant challenge as most micromotors do not provide sufficient disruption forces in in-vivo conditions. In this study, we employed focused scanning laser from conventional confocal microscope to manipulate carbon microbottle based microswimmers. With the increasing of the laser power, the microswimmers' motions translates from autonomous to directional, and finally the high power laser induced the microswimmer explosions, which effectively deliveres microbottle fragments through the cell membrane. It is revealed that photothermally-induced cavitation bubbles enable the propulsion of microbottles in liquids, where the motion direction can be precisely regulated by the scanning orientation of the laser. Furthermore, the membrane penetration ability of the microbottles promised potential applications in drug delivery and cellular injections. As microbottles navigate toward cells, we strategically increase the laser power to trigger their explosion. By loading microswimmers with transfection genes, cytoplasmic transfection can be realized, which is demonstrated by successful gene transfection of GPF in cells. Our findings open new possibilities for cell injection and gene transfection using micromotors.

## Introduction

Micro/nanomotors (MNM), which can effectively convert ambient energy into mechanical energy, is being developed into viable biomedical tool for non-invasive targeted drug delivery and micromanipulation.[1,2] Usually, the self-propulsion of MNMs is achieved through the mechanisms such as self-diffusiophoresis,[5–7] self-electrophoresis,[8–10] and bubble-propelling [11–14], where the chemical reaction on the MNM surface is required. On the other hand, the external field-propelled motors are powered with direct external energy input, such as magnetic,[15–17] electric,[18–20] acoustic,[21–23] light fields,[24–28] or the combination of multiple energy sources.[29–32] The autonomously untethered motions of these MNMs offer promising applications in diverse fields such as environmental remediation,[33–35] sensing,[36–38] micro-molding[39,40] and biomedicine.[41–44]

Particularly, in the application of biomedicine, micromotors are anticipated to navigate noninvasively through the human body, reaching disease-affected regions that traditional medical devices cannot access.[3,45] They are expected to perform specialized tasks, including targeted drug delivery,[46–49] minimally invasive surgery,[50–52] precise nanosurgery[53,54] and detoxification.[55–57] However, only limited propulsion mechanism has been adopted to biomedical nanobot application, which is mainly focused on magnetic field driven and biohybrid nanobot[ref], while only limited nanobots demonstrate sufficient driven force to overcome physiological barriers, such as the blood-brain barrier, blood-testis barrier, mucus, biofouling, and cell membrane barrier.[3,58,59]

On the other hand, extensive researches have focused on developing the optical controlled micro/nanomotors due to the advantage of control flexibility and biocompatibility for therapeutic and diagnostic procedures in biomedicine.[50,62,63] Initially, Jiang et al. proposed that Janus particles can be propelled by a defocused laser beam through self thermophoresis, with the laser power of $10^4$ W cm$^{-2}$.[24] As the high laser density might cause bioactivity loss, recent studies have demonstrated the enhanced phototherapy effectiveness of Janus micromotors with specific structures, such as a spiky head, through both confocal laser scanning and continuous laser irradiation.[47,48,58] Nevertheless, a comprehensive investigation on the operation mode and intensity threshold of lasers to avoid bioactivity damage is lacking. Navigating optical micro motor systems with strong instantaneous penetration force while minimizing adverse effects on organism remain challenging. Recently, it is reported that cavitation bubbles, produced by transformation of light energy into vapor energy,[64–66] can generate high-speed, momen- tary jet flows upon collapsing, which allows them to propel motors and permeate biological barriers.[67–69] Inspired by this, we developed carbon based micromotors which can effectively absorb light energy and generate cavitation bubbles for biomedical applications.

Here, a kind of carbon based micro bottles (CMBs) is prepared as micromotors, and their motion modes under a continuous laser and a scanning laser are investigated systematically. We found that the scanning laser can efficiently propel CMBs at low laser intensity compared to the continuous laser, while inducing a intriguing directional motion mode of CMBs at relatively high light intensity. We investigate the mechanism behind this directional motion both numerically and experimentally. Furthermore, we demonstrate the potential of scanning laser propelled CMBs for targeted cell navigation and their ability to overcome membrane barriers through cavitation bubble induction, thereby enhancing plasmid transfection efficiency. These findings provide valuable insights into the field of laser-propelled micromotors and highlight their potential in revolutionizing various biomedical procedures.

## Methods

### Preparation of materials

**Carbon micro bottles**

The recipe of CMBs were as dipicted by ref.[70] All relevant reagents including Sodium oleate, 98%, Dribose (98%), furfural (99%), and oleic acid (98%) were supplied by Dickman and used without further purification. The micro bottle structure was characterized by scanning electron microscopy (SEM). As shown in Figure 1b, the diameter of the CMB motor is approximately 2 $\mu$m.

**Nanobubble water**

To increase the gas concentration in water, Triton (99%, Sigma) was added into DI water with concentration of 0.5 wt%, followed by processing in an ultrasonic emulsifier with the power of 80% for 0.5, 1, 2, and 5 min. Subsequently, the resulting gas nanobubbles in the water were characterized using a Dynamic Light Scattering (DLS) device.

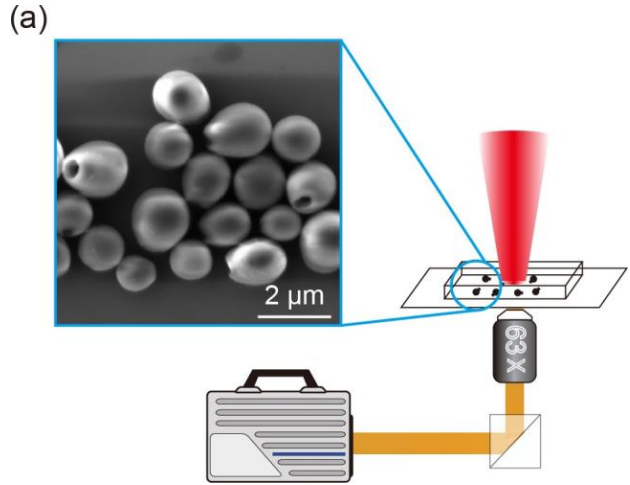

Figure 1: (a) Schematic of the laser setup for CMB propulsion. The inset is the SEM image of the as-prepared CMB motors.

**HeLa cells**

The Hela cell line was obtained from the American Type Culture Collection (ATCC). The cells were cultured at 37 °C in 5% $CO_2$ and maintained in Dulbecco's modified Eagle's medium (DMEM) supplemented with 10% fetal bovine serum (FBS), 100 $\mu$g mL$^{-1}$ penicillin, and 100 $\mu$g mL$^{-1}$ streptomycin. Cells were passaged twice a week using trypsin-EDTA solution (all from Invitrogen).

**Setup**

A confocal microscope (Leica sp8, Germany) with 4 laser beams with wavelengths of 488, 532, 762 and 1064 nm was employed to generate scanning laser. The sketch of the setup is shown in Fig. 1. In this system, laser irradiation is from the top and the optical images are captured by a optical camera with a 63 × objective from the bottom of the CMBs sample. The CMBs are immersed in an aqueous phase and held within a glass tube sealed at both ends with wax. In a typical experiment, all four laser beams were switched on for driving

CMBs and optical imaging simultaneously. For the continuous laser setup, a high power continuous supercontinuum laser source (SC-PRO, YSL Photonics) was employed and an inverted microscope was used for the imaging. Conversely, for the pulse laser experiment, a pulse laser generator (Spectra Physics Quanta Ray 1064 nm) was employed alongside an upright microscope for imaging. Throughout all experiments, a laser power meter (Sper Scientific 840011) was used to measure the intensity of the lasers.

## Fluorescent labeling test for overcoming membrane barriers

A fluorescent labeling test was performed to demonstrate the ability of cavitation bubbles induced by CMBs to overcome membrane barriers of Hela cells. In this domstration, the CMBs were labeled by green fluorescent protein (GFP, 99%, Sigma-Aldrich) in green. Cytoskeleton were labeled by filaggrin (99%, Sigma-Aldrich) in red and cell nucleus were labeled by 4',6-diamidino-2-phenylindole (DAPI, 99%, Sigma-Aldrich) in blue. A confocal microscope was employed for the three-dimensional reconstruction of Hela cells subsequent to CMBs detonation. Laser beams with wavelengths of 488, 532, and 762 nm were employed to excite DAPI, GFP, and filaggrin correspondingly.

## Plasmid transfection experiment

The enhanced green fluorescent protein (EGFP) plasmid was employed to investigate cell transfection by CMBs. Prior to the experiment, Hela cells were cultured until they reached 80% confluency. Subsequently, they were seeded onto a glass-bottom dish and incubated overnight. Following this, DMEM medium was prepared to support the cultured cells. Triton X-100 (Thermo Fisher Scientific) was added in the DMEM medium with the concentration of 0.5%, and subjected to ultrasound for 5 minutes before being supplemented with CNBs and EGFP. Afterward, the cultured cells were washed three times with phosphate-buffered saline (PBS, Invitrogen), followed by immersion in the prepared DMEM medium. The expression of EGFP in Hela cells driven by CMBs was visualized using confocal laser scanning microscopy.

# Numerical model of interactions between a cavitation bubble and a CMB/a cell membrane

We endeavored to simulate the dynamics of cavitation bubbles under diverse boundary conditions using a Boundary Integral Method (BIM). Herein, we provide a concise overview of this methodology. Firstly, we estimate the associated Reynolds number (Re), defined as $Re = UR_m/\nu$, where $U$ represents the characteristic velocity, $R_m$ signifies the maximum radius the bubble, and $\nu$ denotes the kinematic viscosity of the fluid. For micron-sized bubbles, Re falls within the range of $O(10\text{-}10^2)$. Additionally, we calculate the Weber number (We), defined as $We = \rho U^2 R_m/\sigma$, where $\rho$ signifies the fluid density and $\sigma$ represents the coefficient of surface tension. Typically, We ranges between $O(1\text{-}10)$. Consequently, we employ a combination of viscous potential flow theory and BIM to simulate the dynamic behaviors of cavitation bubbles. The flow field surrounding the bubble is governed by the Laplace equation

$$\nabla^2 \varphi = 0, \qquad (1)$$

where $\varphi$ represents the velocity potential, and the velocity field $\boldsymbol{u}$ can be expressed as the gradient of $\varphi$, i.e., $\boldsymbol{u} = \nabla \varphi$. Instead of directly solving the Laplace equation, we resolve the boundary integral equation in numerical simulation. It is widely acknowledged that BIM reduces the dimensionality by one; only the boundaries of the flow field need to be considered and meshed into grids. For the sake of simplicity, we opt for the axisymmetric BIM in this investigation owing to the geometric characteristics involved.

The influence of viscosity within the boundary layers can be addressed through dynamic boundary conditions applied to the bubble surface, expressed as follows:

$$\frac{d\varphi}{dt} = \frac{P_\infty - P_b}{\rho} + |\nabla \varphi|^2 \frac{1}{2} - \frac{\sigma \kappa}{\rho} - 2\nu \frac{\partial^2 \varphi}{\partial n^2}, \qquad (2)$$

where $P_\infty$ is the ambient hydrostatic pressure, $P_b$ the gas pressure inside the bubble, $\rho$

the density of the fluid. The last two terms on the right side present the surface tension and viscous corrections, respectively. This comprehensive boundary condition is crucial for accurately simulating the behavior of cavitation bubbles in viscous fluids and is incorporated into the computational model to capture the complex dynamics of bubble formation, growth, and collapse under various conditions. The gas pressure inside the bubble is calculated from the ideal gas equation. To update the velocity potential $\varphi$ on the bubble surface in the time domain, Equation 2 is used, and numerical methods such as the fourth-order Runge-Kutta method are employed for enhanced accuracy in the simulation process.

The kinematic boundary conditions on the bubble surface is expressed as:

$$\frac{\mathrm{d}r}{\mathrm{d}t} = \nabla \varphi, \tag{3}$$

which is essential for updating the location of the bubble surface over time, accounting for the movement of the bubble surface itself.

The boundary condition for a wetted particle surface or rigid wall is expressed as follows:

$$\frac{\partial \varphi}{\partial n} = \boldsymbol{U} \cdot \boldsymbol{n}, \tag{4}$$

where $U$ is the velocity of the object (e.g., the translational velocity of the particle), $\boldsymbol{n}$ denotes the outward unit normal vector on the boundary surface. This boundary condition ensures that no fluid penetrates the boundary surface.

The first numerical setup is about the jetting behavior of a cavitation bubble near a cell that attached to the rigid wall. Due to the relatively flat morphology of cells adhering to the basal wall surface, as well as the significantly higher fluid viscosity inside the cells compared to water, we will employ a bubble model near rigid boundaries to assess and analyze the jet velocity and impact pressure of the bubble. We employ wall Green's functions and the method of images to avoid directly discretizing the wall boundary, thus simulating the dynamic behavior of bubbles near the wall. This approach allows us to conserve significant computational resources.

In the second numerical setup for the bubble-particle interaction, we utilize an auxiliary function method to decouple the calculation of the hydrodynamic force and acceleration driven by the bubble. This approach not only accounts for bubble dynamic behaviors but also considers particle acceleration. As particle motion encounters fluid viscous resistance, we have included this factor in our computational model as well. For the determination of the drag coefficient of particulate matter, interested readers can refer to Borkent's work. More details about the numerical models can be referenced from our previous research.

## Results and discussion

### Light intensity dependent motion behaviors

The motions of catbon micro bottles in water under scanning laser, continuous laser and pulse laser have been monitored with an optical camera. Figures 2a1-3 show three typical moving trajectories of CMBs under the scanning laser irradiation with intensities of 2.0, 10.0 and 20.0 W cm$^{-2}$, respectively. Under an intensity of 0.2 W cm$^{-2}$, the CMB shows the Brownian motion around a fix point with the velocity of around 3 $\mu$m s$^{-1}$ (Fig. 2a1 and c). When the CMBs were exposed to the scanning laser intensity of 10.0 W cm$^{-2}$, they exhibit a nondirectional autonomous motion with the average velocity of approximately 10 $\mu$m s$^{-1}$ (Fig. 2a2 and c), which has been revealed that the ejection of the heated fluid from the open neck of the CMB leads to its self-propelling.[71] Intriguingly, the CMB exhibits a fast shooting behavior (Fig.2a3), with the maximum instantaneous velocity of 130 $\mu$m s$^{-1}$ (Fig.2c, left panel). The mean square displacement (MSD) shown in Fig. 2d also indicates that the directional motion of CMBs corresponds to significant super diffusion.

In contrast, the autonomous motion of the CMBs under continuous laser irradiation requires a significantly high laser power. The CMBs demonstrate evident autonomous motion with a maximum velocity of approximately 13 $\mu$m s$^{-1}$ (Fig. 2c, right panel) when the laser power reaches 3000 W cm$^{-2}$ (Fig. 2b), enabling them to effectively overcome thermal fluctuations. When the laser power ranges from 0.5 to 2500 W cm$^{-2}$, thermal fluctuations dominate the motion of CMBs, resulting in a velocity of approximately 6 to 8 μm s$^{-1}$ (Fig.2e). Additionally, a pulse laser irradiation is also utilized to propel the CMBs, where only an intensity of 0.05 W cm$^{-2}$ is sufficient to initiate the autonomous motion of CMBs (Fig. S1 and Movie S5 in the Supplementary Materials).

Compared to continuous laser irradiation ($P_\ell$ = 3000 W cm$^{-2}$), both scanning laser ($P_\ell$ = 10 W cm$^{-2}$) and pulse laser ($P_\ell$ = 0.05 W cm$^{-2}$) can propel the micro motors with relatively low laser intensities, as they compress the illumination in space and time. It is important to note that instantaneous power refers to the power of the laser during the time it is deposited on the targets. For a pulse laser with a duration of 6 ns and frame rate of 10 Hz, an average power of 0.01 W cm$^{-2}$ corresponds to an instantaneous power of approximately 10$^4$ W cm$^{-2}$. Meanwhile, the instantaneous power of a scanning laser depends on the scanning rate and target objects. For example, at a laser power of 20 W cm$^{-2}$ and a scanning rate of 7.7 Hz, approximately 10000 W cm$^{-2}$ is instantaneously deposited on CMBs (see Supplementary Materials for more details).

To get more insight into the intriguing directional fast motion behaviors of CMBs under a scanning laser, we quantitatively investigate the dependence of the motion modes of CMBs on control parameters, namely laser intensity and scanning rate. The laser power was adjusted in a range of 0 to 20 W cm$^{-2}$ while varying the scanning rates at 7.7 Hz, 3.8 Hz and 1.9 Hz. In order to quantify motion characteristics of CMBs, we employed the maximum velocity of CMBs as a representative parameter. As shown in Fig. 2f, for a scanning rate of 7.7 Hz, increasing the light intensity from 0.2 to 20 W cm$^{-2}$ results in an increase in the maximum velocity $V_{max}$ of CMBs from less than 5 μm s$^{-1}$ to approximately 130 μm s$^{-1}$. The motion behavior depends on light intensity and can be categorized into three types accordingly. The range of light intensity between 0.4 and 4.0 W cm$^{-2}$ corresponds to the Brownian movement, whereas the range of light intensity between 6.0 and 14.0 W cm$^{-2}$ corresponds to autonomous motion. While the light intensity is in the range of 16.0 to 20.0 W cm$^{-2}$, CMBs start to demonstrate directional fast motion behaviors. Additionally, the scanning rate of confocal microscope also significantly affects the motion modes of CMBs (Fig.2f). At a scanning rate of 3.8 Hz and 1.9 Hz, the minimum laser power required to trigger directional motion decreases to 12 and 8 W cm$^{-2}$ respectively (Fig. 2f).

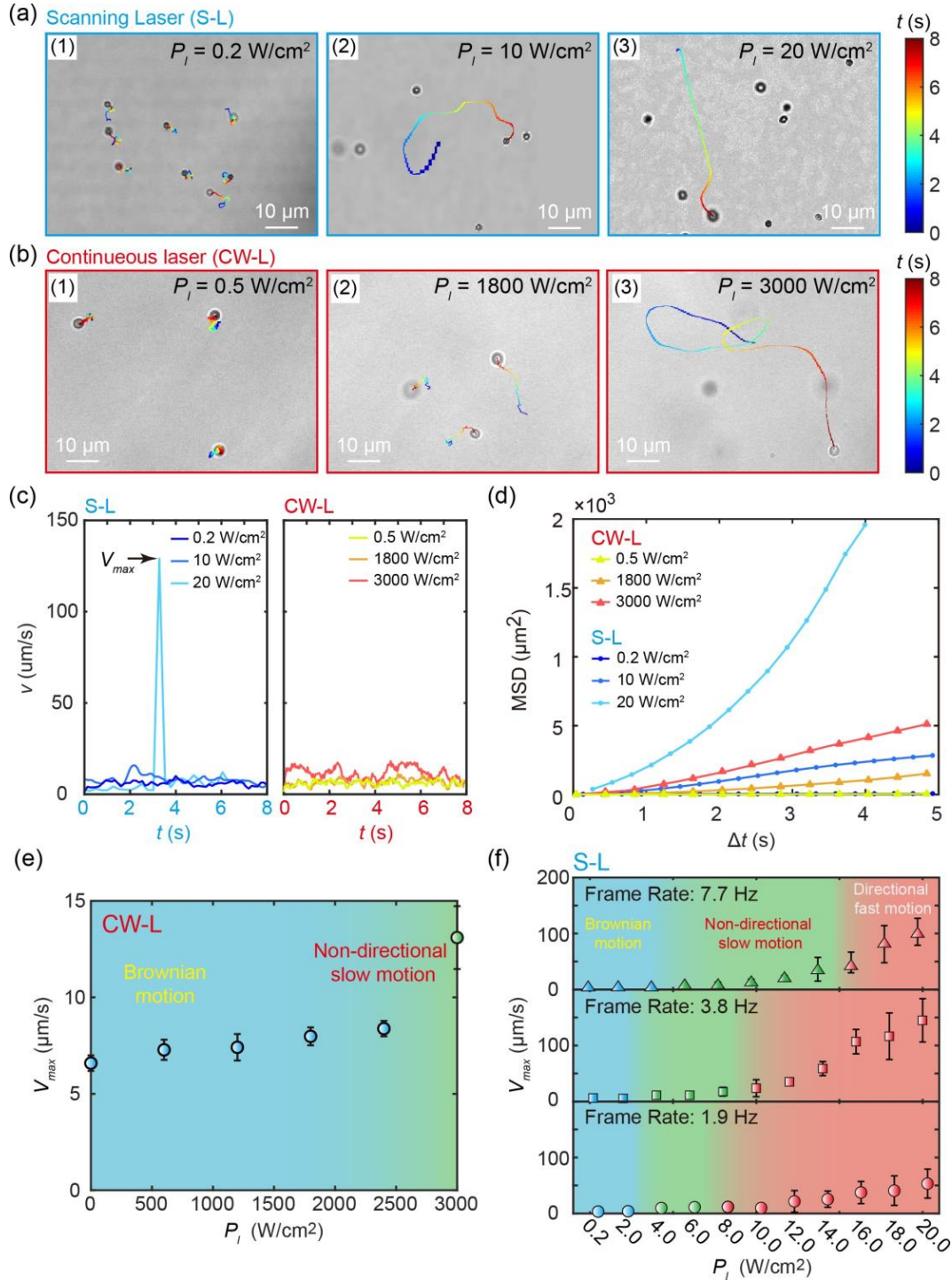

Figure 2: (a) Trajectory images of CMB under scanning laser irradiation with powers of 0.2, 10 and 20 W cm$^{-2}$. (b) Trajectory images of CMB under continuous irradiation with powers of 0.5, 1800 and 3000 W cm$^{-2}$. (c) Instantaneous velocity of CMB motors under a scanning laser (left panel) at intensities of 0.2, 10, and 20 W cm$^{-2}$, as well as a continuous laser (right panel) at intensities of 0.5, 1800, and 3000 W cm$^{-2}$ for a duration of 8 s. (d) Mean square displacement (MSD) of CMBs for scanning laser and continuous laser for various intensities. (e) Maximum velocity distribution of CMBs for continuous laser intensities from 0 to 3000 W cm$^{-2}$. (f) Maximum velocity distribution of CMBs with laser scanning rate of 7.7, 3.8 and 1.9 Hz.

## Driving mechanism of micro bottles

To understand the origin of the high-speed directional motion of CMBs induced by a scanning laser, we reduced the scanning speed to decelerate this rapid phenomenon. Figure 3a displays the dynamic state of a CMB with laser scanning speed decreased to 100 lines s$^{-1}$. It is revealed that only the upper portion of the CMB is exposed to laser irradiation, as well as the top section continued to appear below its former location. This suggests that as the laser scans across the top section of a CMB, resulting the CMB to be ejected downwards at a velocity greater than that of the scanning laser.

Given the ability of CMBs to convert light energy into heat, it is crucial for us to determine the temperature increase on CMBs under laser scanning. Here we numerically estimated the temperature distribution in the vicinity of the CMB when a laser scans through its top portion. It should be noted that the directional motion of the CMB during high-intensity laser scanning is independent of the orientation of CMB's opening neck. Therefore, we focused on a representative state where the opening neck was oriented towards the right. Figure 3b illustrates the temperature distribution field surrounding a CMB under an average power of 8 W cm$^{-2}$, corresponding to an instantaneous power value of 15600 W cm$^{-2}$ within a duration of 2 ms. The highest recorded temperature was found at the top point on the CMB, reaching up to 170 °C.

The temperature increase may induce a phase transition in the surrounding medium. The phase diagram of water, shown in Figure 3c, illustrates the maximum attainable temperature for liquid water under a given pressure, known as the spinodal temperature (represented by the green curve in Fig. 3c). At this temperature, spontaneous homogeneous nucleation of vapor bubbles occurs. However, in most cases, bubble nucleation takes place at temperatures significantly lower than the spinodal temperature due to the presence of tiny cracks, cavities, or pits filled with gas, impurities, or aggregations of gas molecules that can serve as nucleation centers.[72,73] This latter process is referred to as heterogeneous nucleation (indicated by the blue curve in Fig. 3c).

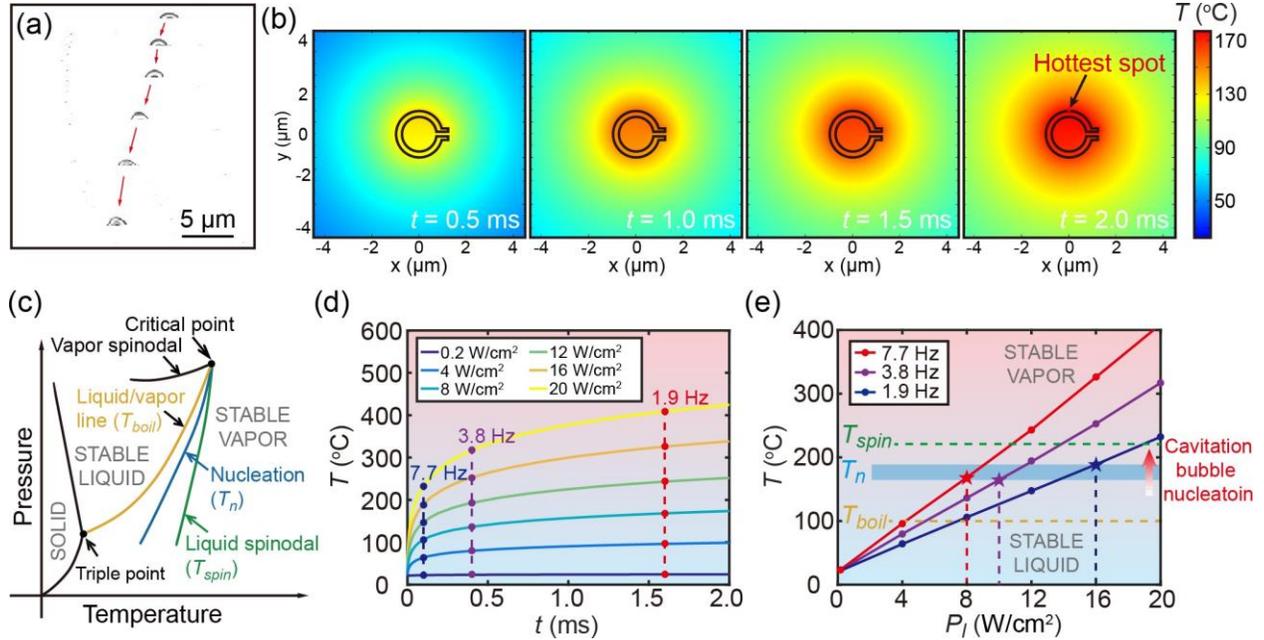

Figure 3: (a) Optical image from confocal microscope at the scaning speed of 200 line/s. (b) Numerical temperature field around a CMB under laser scanning through a CMB under the average laser power of 20 W cm$^{-2}$. (c) Schematics of phase diagram of water. (d) Evolution of maximum temperature on a CMB in time for various laser irradiation. (e) Maximum temperature on a CMB as a function of laser power with scanning rates of 7.7, 3.8 and 1.9 Hz.

Figure 3d displays the numerical results of temporal evolution of temperature at the hottest spot on the CMB for various average laser powers, ranging from 0.2 to 20 W cm$^{-2}$. It should be noted that the duration of laser scans on the CMB is not constant and depends on the scanning rate. In experiments, we can estimate the duration $T_{dur}$ of laser irradiation on the top portion of bottles based on their size and applied scanning rate. As illustrated in Fig. 3d, $T_{dur}$ are 100, 400 and 1600 μs for the scanning frame rate of 7.7, 3.8 and 1.9

Hz, respectively. The relationship between temperature $T$ and laser power $P_\ell$ for these three different scanning frame rates is depicted in Figure 3e which demonstrates a linear dependence between maximum temperature and laser power as it increases from 0.2 to 20 W cm$^{-2}$.

Additionally, as illustrated in Fig. 2f, fast directional motion of CMBs was induced at $P_\ell$ = 8, 10 and 16 W cm$^{-2}$ for the scanning frame rates of 7.7, 3.8 and 1.9 Hz, respectively. Accordingly, the theoretical temperature corresponding to different scanning rates can be derived from the results in Fig. 3e, which are denoted by the red, purple and blue stars, respectively. By referring to the phase diagram depicted in Fig. 3c and considering the theoretical temperature values obtained, it is observed that the nucleation temperature $T_n$, associated with phase transition, falls within the range of boiling temperature and spinodal temperature regions. This finding aligns with previous research outcomes.[74]

Here we note that cavitation bubbles have a lifespan of only a few microseconds, which significantly surpasses the time resolution capabilities of confocal microscopes. To illustrate cavitation bubble nucleation, we conducted an experiment using CMBs within a hydrogel composed of 0.22% hyaluronic acid and 0.11% agar. This mixture has higher viscosity than water, effectively prolonging the life cycle of cavitation bubbles. As displayed in Movie S6, one can observe the formation and subsequent collapse of microbubbles measuring approximately 5 $\mu$m in diameter on top of the CMBs during their rapid directional movement.

## Numerical simulation of the interaction between a cavitation bubble and a micro motor

Here we conduct a numerical simulation to further reveal the mechanism by which cavitation bubbles induce high-speed directional motion of a micro bottle. To simplify the calculation, a micro particle is used instead of a micro bottle. Figure 4b displays the pressure contours during the interaction between a cavitation bubble and a micro particle. The initial time of bubble expansion, as deduced from simulations, is shown in Figure 4b(1-3), while subsequent collapse states are depicted in Figure 4b(4-8). Furthermore, the time-dependent pressure intensity exerted on the CMB was extracted (Figure 4c). The pressure-time curves exhibit an initial peak value of approximately 2 MPa, indicating a propelling force $F_p$ (Figure 4a-1) exerted on CMB during the rapid growth of the bubble, followed by a secondary weak peak during disconnection between the bubble and the particle, corresponding to a relatively weak $F_p$ (Figure 4a-4). Thus, as depicted in Figure 4a, the propulsion of the micro particle is primarily driven by the rapid expansion of the cavitation bubble, followed by a depletion of kinematic energy due to Stokes drag $F_d$ after bubble collapse.

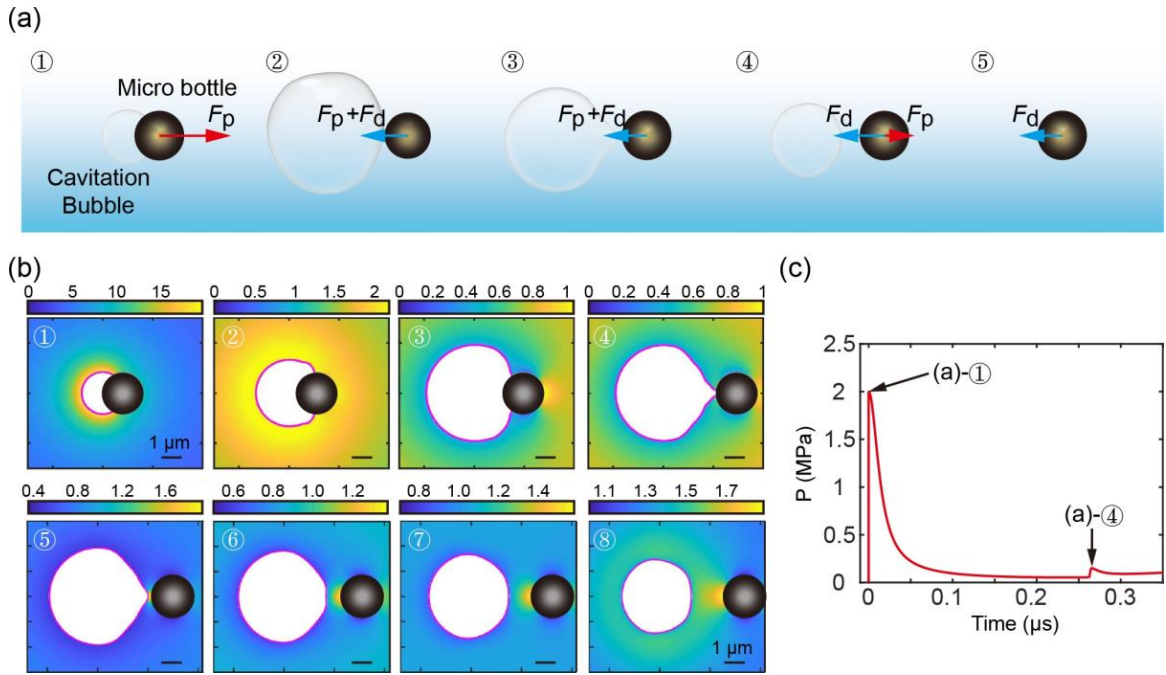

Figure 4: (a) Numerical model of a cavitation bubble with a diameter of 5 um growing on a CMB motor with a diameter of 2 um. (b) The interaction and pressure contours in the numerical simulation. (c) The temporal evolution of the pressure intensity acting on the CMB motor generated by the bubble cavitation.

# Enhancement of cavitation bubbles nucleation by introducing nanobubbles

To optimize the use of light induced cavitation bubbles in biomedical fields, it is essential to lower the required laser intensity. Prior studies have demonstrated that introducing contaminants or gas molecules into water can achieve this effect.[72,73] Here we aim to enhance the efficiency of cavitation bubble generation by incorporating nanobubbles into Triton-infused water through a robust ultrasonic process, thereby reducing the laser power threshold required. Figure 5a displays the dynamic light scattering (DLS) images of nanobubbles following the ultrasonic process at various processing times. As processing time increases from 0 to 5 minutes, more nanobubbles form in the water. The absolute amounts of nanobubbles per volume increases to approximate $4\times10^7$ counts mL$^{-1}$ when the processing time $t_{process}$ increase to 5 min (Figure 5b). The probability density function (PDF) of the nanobubble volume, as depicted in Figure 5b inset, demonstrates an increased uniformity in bubble size after 5 minutes of processing. The resulting nanobubbles predominantly exhibit a diameter around 200 nm.

Subsequently, a series of systematic experiments were conducted by varying the laser power $P_\ell$ from 0.2 to 20 W cm$^{-2}$ with nanobubble concentrations ranging from 0 to $4\times10^7$ counts mL$^{-1}$, while maintaining a frame rate of 7.7 Hz. In the phase diagram depicted in Fig. 5c, an increase in nanobubble concentration $c_{bubble}$ from 0 to $4\times10^7$ counts mL$^{-1}$ resulted in a decrease in the minimum laser power required for inducing directional motion of CMBs, reducing it from 16 to 12 W cm$^{-2}$. Meanwhile, the increase in $c_{bubble}$ leads to a slight increase in the maximum velocity of the CMB. This suggests that nanobubbles present in water serve as nucleation sites and promote the formation of cavitation bubbles.

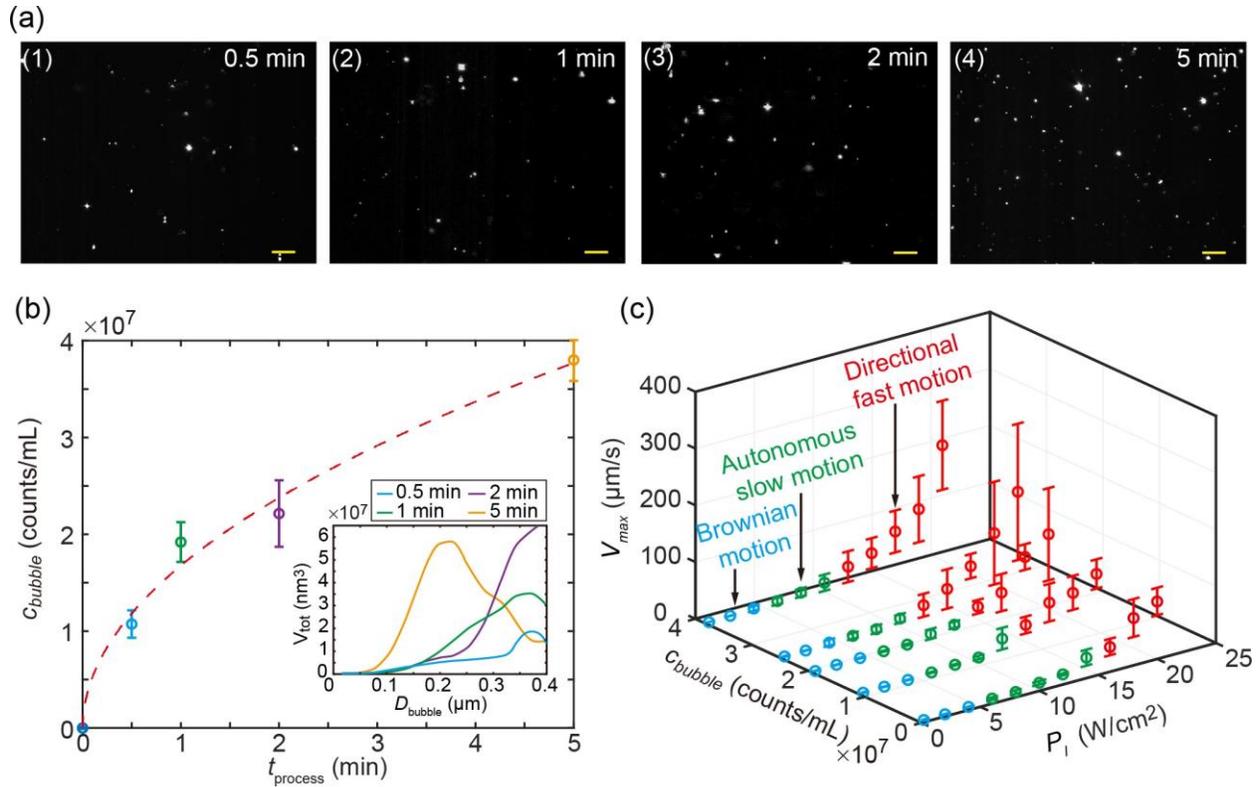

Figure 5: (a) Dynamic light scattering (DLS) images of nanobubbles in water under ultrasonic with 0.5, 1, 2 and 5 min. (b) Volume distribution of bulk nanobubbles for ultrasoic processing time of 0.5, 1, 2 and 5 min. The dash curve was drawn to guide the eye. (c) Maximum velocity and phase diagram of CMBs under various nanobubble concentrations and average laser intensities.

## Selective plasmid transfection in cells based on CMBs

Given the instantaneous release of substantial energy of cavitation bubbles and their resulting directional motion behavior of CMBs, it is possible to employ these CMBs for overcoming biological barriers. In this study, we present a demonstration of using CMBs for achieving selective gene transfection. The strategy and corresponding optical images are depicted in Figure 6a, illustrating the precise navigation and controlled explosion of CMBs around a Hela cell. Initially, cells and CMBs are imaged and located at low light intensities (Fig. 6a-1), minimizing the impact of Brownian motion on their positions. Subsequently, by adjusting the scanning orientation of the laser and increasing its power, cavitation bubble generation is initiated to propel the CMBs towards the cell (Fig. 6a-2). Upon CMBs reaching the cell, an explosion occurred as a result of the nucleation of a cavitation bubble within the CMB (Fig. 6a-3), followed by subsequent collapse of this bubble (Fig. 6a-4). This controlled explosion can be effectively utilized for efficient penetration through cellular membrane barriers.

Numerical and experimental methods are employed to demonstrate the capability of CMBs in overcoming the cellular membrane barrier. Figure 6b illustrates the pressure contours of a cavitation bubble interacting with a cell membrane, which is depicted as a solid boundary due to the Hela cell's viscosity being $10^4$ times higher than water.[75] Upon achieving its maximum size, the cavitation bubble collapses, producing a jet flow directed towards the membrane. The maximum instantaneous pressure intensity on the membrane can reach 8 bar (Figure 6c), equivalent to 2 mN, which is three orders of magnitude more potent than the force necessary for mechanically opening the cell membrane.[61] Moreover, a confirmatory experiment was carried out using fluorescence labeling, with CMBs marked in green, cytoskeleton in red, and cell nucleus in blue. The 3D confocal image in Fig. 6d further indicates that substances transported by CMBs can be effectively delivered into cells.

Given the ability of CMBs to break through cellular membranes, we propose their potential application as a gene delivery tool similar to a gene gun for efficient transfer of plasmids into target cells. Plasmids, which are small extrachromosomal circular DNA molecules, are typically introduced into cells through endocytosis. However, the process of endocytosis lacks selectivity and is time-consuming, with an average duration of approximately 6 hours.

Cavitation bubbles induced by CMBs can effectively address the aforementioned issues, enabling rapid and targeted plasmid transfection in cells. Figure 6e illustrates a series of fluorescent images capturing the progression of cell fluorescence following bubble explosion. The intensity of fluorescence emitted by the cells gradually increases over a period of 360 minutes. Figure 6f presents the temporal evolution of fluorescent intensities for varying explosion times, indicating that plasmid expression within cells initiates at 30 minutes and reaches its maximum after approximately 120 minutes, exhibiting an upward trend as explosion times increase. Moreover, we successfully demonstrate the selective targeting of specific cells using CMBs and achieve efficient plasmid transfection, as evidenced by the formation of 'HKU' letters through fluorescence imaging. This highlights the potential application of CMB-based scanning lasers for precise plasmid transfection.

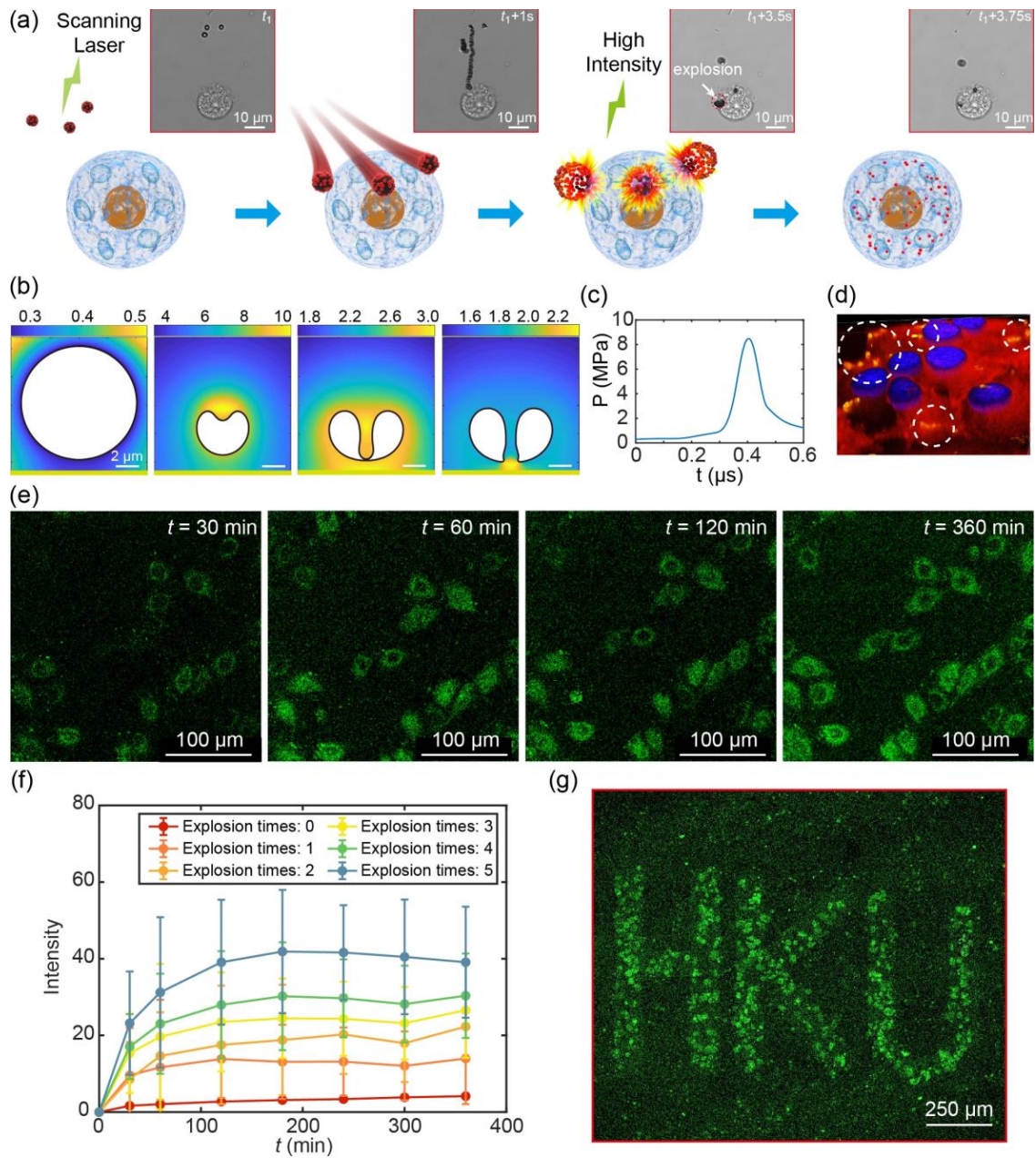

Figure 6: (a) Sketches and successive optical images of CMB navigating toward a cell and subsequent explosion. (b) Numerical simulation of explosion of a cavitation bubble and the induced jet flow near a cell membrane. (c) The temporal evolution of the pressure intensity acting on the cell membrane. (d) 3D constructed fluorescence image of the cell after explosions by CMBs induced cavitation bubbles. CMBs, cytoskeleton and nucleus are labeled by green, red and blue, respectively. (e) Fluorescence images of cells after explosions by cavitation bubbles. (f) Fluorescence intensity of cells as a function of time for various explosion times. (g) 'HKU' letters writing of cells from plasmid transfection by selective explosion.

## Conclusion

In conclusion, this study experimentally and numerically demonstrates the potential of carbon-based micromotors in the field of biomedicine, particularly for targeted drug delivery and precise nanosurgery applications. By utilizing cavitation bubbles' unique properties and optimizing their generation through a scanning laser from a confocal microscope, we achieved directional motion and manipulation of microbottles at relatively low power intensity. This enables us to navigate micromotors by adjusting the orientation of the scanning laser. Furthermore, we showcased the practical application of this approach in intracellular plasmid transfection by guiding microbottles towards target cells and generating a jet capable of penetrating cell barriers. This accomplishment not only validates the effectiveness of our method but also highlights the potential of micromotors in revolutionizing various biomedical procedures.

## Acknowledgement

This work was supported in part by the Hong Kong Research Grants Council (RGC), the Collaborative Research Fund (C7082-21G), RGC Research Fellowship (RFS2122-7S06), Croucher Foundation Senior Research Fellowship (2022), Hong Kong Quantum AI Lab, AIRInnoHK of the Hong Kong Government, and Seed Fund for Strategic Interdisciplinary Research Scheme by the University of Hong Kong.

# Supporting Information Available

This will usually read something like: "Experimental procedures and characterization data for all new compounds. The class will automatically add a sentence pointing to the information on-line: